\documentclass{aa}  
\usepackage{xcolor}
\usepackage[varg]{txfonts}
\usepackage[colorlinks=true, allcolors=blue, urlcolor=blue, citecolor=blue, linkcolor=blue]{hyperref}
\usepackage{graphicx}

\newcommand{\kms}{km\,s\ensuremath{^{-1}}\xspace}

\begin{document}

\title{Orbital decomposition of the nuclear regions in the early-type galaxy FCC~47: Unveiling the nuclear cluster origin}
\titlerunning{Orbital decomposition of the nuclear regions in the early-type galaxy FCC~47}

   \author{J. Lamprecht
   \inst{1,2}
   \and
   A. Feldmeier-Krause
   \inst{2} 
      \and
   M. Lyubenova
   \inst{3} 
   \and
   K. Fahrion
      \inst{2}
   \and
   S. Thater
      \inst{2}
   \and
   P. Jethwa
      \inst{2}
   \and
   S. Reiter
   \inst{2}
      \and
      \newline
   J. Falcón-Barroso
      \inst{4,5}
         \and
         T. I. Maindl
      \inst{2,6}
      \and
   G. Santucci
   \inst{7}
         \and
   I. Breda
   \inst{2}
   }

             \institute{Department of Theoretical Physics and Astrophysics, Masaryk University, Kotlářská 267/2, 611 37 Brno, Czech Republic \\
\email{lamprecht@sci.muni.cz}
\and
Department of Astrophysics, University of Vienna, Türkenschanzstraße 17, 1180 Vienna, Austria
\and
European Southern Observatory (ESO), Karl-Schwarzschild-Straße
 2, 85748 Garching, Germany
\and
Instituto de Astrofísica de Canarias, Vía Láctea s/n, E-38205 La Laguna, Tenerife, Spain
\and
Departamento de Astrofísica, Universidad de La Laguna, E-38200 La Laguna, Tenerife, Spain
\and
SDB Science-driven Business Ltd., Faneromenis Avenue 85, Ria Court 46, Suite 301, 6025 Larnaca, Cyprus
\and
CSIRO Space \& Astronomy, PO Box 1130, Bentley, Western Australia 6102, Australia
}

   \date{Received 14 November 2025 / Accepted 24 January 2026}

\abstract
{Nuclear star clusters (NSCs) are among the densest stellar systems in the Universe and often coexist with supermassive black holes (SMBHs) at galaxy centres. While SMBH formation histories are essentially lost, NSCs preserve evolutionary imprints through their stellar populations and stellar kinematics, reflecting the cumulative effects of mergers, accretion, and internal dynamical evolution.}
{We aim to investigate the orbital structure of the unusually large NSC in FCC~47 (NGC~1336) by decomposing its stellar orbits into dynamically distinct components.}
{We extract stellar kinematics, and in particular the line-of-sight velocity distributions (LOSVDs), from VLT/MUSE integral-field spectroscopy using the non-parametric \textsc{Bayes-LOSVD} approach, and apply triaxial Schwarzschild orbit-superposition modelling with the DYNAMITE software. We decompose the orbit library into hot, warm, cold, and counter-rotating components.}
{We detect triple-peaked LOSVDs in the nucleus, indicating a complex orbital structure. The NSC forms a counter-rotating, kinematically decoupled component. A hot pressure-supported component, a warm counter-rotating structure and a counter-rotating cold disk in the centre suggest hierarchical assembly via early star cluster accretion and later in situ star formation.}
{Our orbital decomposition of FCC~47 supports a hybrid formation scenario for this NSC. Dynamically distinct substructures reflect the interplay of accretion and in situ star formation during galaxy evolution.}


\maketitle

\section{Introduction}

Nuclear star clusters (NSCs) are extremely dense and massive star clusters in the central regions of galaxies \citep{Walcher2005, Hopkins2010, Neumayer2020}. They are highly luminous and compact, resulting in a tremendous rise in the surface brightness in the centre of their host galaxy. There is ample evidence that NSCs coexist with supermassive black holes (SMBHs) at galaxy centres (e.g. \citealt{Seth2008, Graham2009, Schoedel2009, Nguyen2018, Nguyen2019}). However, in contrast to SMBHs, NSCs preserve their evolutionary imprints in their stellar populations and kinematics, reflecting cumulative effects of mergers, accretion, and internal dynamical evolution. NSCs have been observed to have strong rotation in external galaxies \citep{Lyubenova2013, Lyubenova2019, Pinna2021} as well as in the Milky Way \citep{Feldmeier2014}. Typical NSC effective radii range from 3–10~pc, comparable to globular clusters (GCs), and their masses span $10^5$–$10^8\,M_\odot$, occasionally reaching up to $10^9\,M_\odot$ in the most massive hosts \citep{Spengler2017,Neumayer2020, Hoyer2021}. They are commonly found in galaxies with stellar masses between $10^8$ and $10^{10}\,M_\odot$, and more than 70\% of galaxies in this mass range have an NSC. The nucleation fraction reaches a peak value of 90\% for galaxies with stellar masses of about $10^{10}\,M_\odot$ and declines for both higher and lower masses \citep{SanchezJanssen2019, Hoyer2021, Zanatta2024}. Despite their ubiquity, the formation and evolutionary pathways of NSCs remain poorly understood. Traditionally, NSC formation is understood through two primary channels: in situ star formation and the accretion of GCs or external stellar systems, see the review by \cite{Neumayer2020}. In the in situ scenario, the NSC might form independently of the GC system, with gas infall from the surrounding region, such as accretion from a disk, driving efficient star formation and rapid metal enrichment \citep{Bekki2007, Antonini2015,Neumayer2020}. This process typically results in NSCs with strong rotation, high metallicities, and recent star formation as observed in many late-type galaxies \citep{Milosavljevi2004,Seth2006,Feldmeier-Krause2015, Feldmeier-Krause2017, Kacharov2018, Fahrion2024}. On the other hand, the dry GC accretion scenario involves the inspiral of GCs or external stellar systems due to dynamical friction, leading to NSCs that reflect the metallicity of the accreted GCs \citep{Tremaine1975, Capuzzo-Dolcetta1993,Capuzzo-Dolcetta2008, Fahrion2021b}. While this scenario often predicts weaker rotation due to the random infall directions of GCs, simulations have shown that NSCs formed through mergers of GCs can still have significant rotation if they keep the angular momentum of their formation origin \citep{Hartmann2011, Tsatsi2017, Lyubenova2019}. \cite{Guillard2016} have shown that these processes may operate together in a hybrid scenario of gas-rich proto-star cluster accretion. The NSC in the Milky Way may represent a prominent case where both proposed formation mechanisms contribute. Its star formation history and the presence of young stars point towards in situ formation, while the recent discovery of a metal-poor stellar population could be the result of an infalling, metal-poor GC \citep{Feldmeier-Krause2020, Arca2020}.

Given these results, it is desirable to disentangle dynamically distinct components to constrain NSC formation history. While dynamical modelling has been used multiple times to determine the masses of both NSCs and SMBHs \citep[e.g.][]{Walcher2005,Seth2010,Neumayer2012,Feldmeier-Krause2017,Thater2023}, so far, orbital distributions of NSCs have not yet been explored in depth in external galaxies. Stellar orbit distributions, particularly their angular momentum properties, offer essential clues to the dynamical assembly history and potential kinematic substructures, such as kinematically decoupled cores (KDCs), counter-rotating components, or embedded disks.

Studying NSCs in external galaxies remains observationally challenging due to their small sizes. However, FCC~47, an early-type galaxy in the Fornax Cluster, hosts a remarkably large NSC ($R_{\text{eff,NSC}} = 66.5 \pm 11.1$\,pc, see Table~\ref{tab:properties}), allowing it to be formally resolved in ground-based integral-field data taken with the Multi Unit Spectroscopic Explorer (MUSE, \citealt{Bacon2010}) assisted by adaptive optics under the science programme 60.A-9192 (PI: Fahrion). The MUSE point spread function (PSF) of $0.7$\arcsec at full width half maximum (FWHM) is comparable to the NSC's effective radius, enabling spatially resolved kinematic analysis. FCC~47 also exhibits a rich GC system and a complex star formation history, making it an ideal target to investigate NSC formation mechanisms. \cite{Fahrion2019} obtained maps for the stellar kinematics and stellar population properties of FCC~47 whilst characterising the younger stellar body, the NSC, and the rich GC system simultaneously. This analysis revealed that the NSC is kinematically distinct, overmassive, and counter-rotating with respect to the host galaxy, with a rotation axis nearly orthogonal to that of the main stellar body. \citet{Thater2023} later combined the same VLT/MUSE dataset with high-resolution adaptive-optics-assisted Spectrograph for INtegral Field Observations in the Near Infrared (VLT/SINFONI, \citealt{Eisenhauer2003}) data of FCC~47 with dynamical modelling, incorporating a spatially varying stellar mass-to-light ratio ($M_*/L$) that takes changes of the stellar populations and initial mass function (IMF) into account. They measured a BH mass of $M_{\mathrm{BH}} = 4.4^{+1.2}_{-2.1} \times 10^{7} M_{\odot}$ using $3\sigma$ confidence intervals and verified strong rotation and high angular momentum of the NSC, initially measured by \cite{Lyubenova2019} and confirmed by \cite{Fahrion2019}. More generally, \cite{Fahrion2021} studied NSC formation pathways and provided a quantitative determination of the relative strength of the two main formation processes in the build-up of individual NSCs. For FCC~47, the NSC mass of \( M_{\text{NSC}} = 7.3 \pm 1.2 \times 10^8 \, M_\odot \) places it firmly in the regime where in situ formation is expected to dominate. However, since the NSC in FCC~47 is counter-rotating and appears kinematically decoupled from the main galaxy body, this suggests that GC accretion or the infall of an external stellar system may have contributed to its assembly, potentially explaining the preserved opposite sense of rotation. Alternatively, the observed kinematic decoupling may not directly trace the NSC formation pathway itself, but could instead be the result of a major merger that has significantly altered the dynamical structure of FCC~47.

Building on these results, we performed a full orbital decomposition of FCC~47 using Schwarzschild orbit-superposition modelling with the DYNAMITE software \citep{dynamite, Thater2022} using the same VLT/MUSE data as in \cite{Thater2023} and \cite{Fahrion2019}. In contrast to earlier approaches, we used line-of-sight velocity distributions (LOSVDs) extracted with the \textsc{Bayes-LOSVD} code \citep{Falcon-Barroso2020}, which is particularly advantageous for systems that consist of multiple dynamical components, such as an NSC embedded in a host galaxy. This is the first application of the \textsc{Bayes-LOSVD} code to an NSC. We constructed dynamical models and classified orbits based on their angular momentum. We identified cold, warm, hot, and counter-rotating components, and connected these structures to unveil the NSC's formation history.

This paper is structured as follows: In Sect.~\ref{section2} we present the photometric and spectroscopic data, including the MUSE and HST datasets. In Sect.~\ref{section3} we detail the kinematic extraction based on the MUSE data with the \textsc{Bayes-LOSVD} approach. In Sect.~\ref{section4} we describe the dynamical modelling approach and the construction of orbit libraries. In Sect.~\ref{section5} we discuss the results of the orbital decomposition, and in Sect.~\ref{section6} we summarise our findings and discuss their implications for NSC formation scenarios. We conclude in Sect.~\ref{section7}.

\begin{table}[]
    \centering
     \caption{Basic properties of FCC~47 and its NSC. }   \begin{tabular}{p{4cm}p{2.5cm}p{1cm}}
    \hline \hline
    \noalign{\smallskip}
        FCC~47 & & Notes  \\
        \noalign{\smallskip}\hline\noalign{\smallskip}
        \raggedright Morphological type & S0 & 1\\
        Distance [Mpc] & 18.3 $\pm$ 0.6 & 2\\
        Physical scale [pc arcsec$^{-1}$] & 88.72 & \\
        Inclination [\degr] & 47 $\pm$ ~4 & 3 \\
        Effective radius [kpc] & 2.40 & 4 \\
        Galaxy stellar mass [$M_{\odot}$] & (1.0 $\pm$ ~0.1) $\times 10^{10}$ & 5 \\
        $\sigma_{cen}$ [km s$^{-1}$] & 106 $\pm$ 5 & 6\\
       \noalign{\smallskip} \hline \hline\noalign{\smallskip}
        NSC & & Notes  \\
        \noalign{\smallskip}
        \hline
        \noalign{\smallskip}
        \raggedright Effective radius [pc] &  $66.5 \pm 11.1 $ & \\
        Effective radius [$\arcsec$] & $ 0.75 \pm 0.125$ & 7\\
        NSC mass [$M_{\odot}$] & $7.3 \pm 1.2 \times 10^{8} $& 8\\
        \noalign{\smallskip}
        \hline

    \end{tabular}
\tablefoot{(1) \cite{deVaucouleurs1991}. (2)  \cite{Blakeslee2009}. (3) \cite{Thater2023}. (4) derived by applying the mge\_half\_light\_radius routine of the JamPy package on the MGE of the stellar mass model (Section~\ref{sec:photometricdata}) \citep{Cappellari2002, Cappellari2008}. (5) \cite{Thater2023} using a dynamical Schwarzschild orbit-superposition model with varying stellar populations and varying IMF. (6) \cite{Thater2023}; (7) \cite{Turner2012}, measured in the F475W Filter ($\approx g-$band); (8) \cite{Fahrion2019} photometric mass using stellar population models.}
    \label{tab:properties}
\end{table}

\section{Data}\label{section2}

\subsection{HST photometric data and stellar mass model}\label{sec:photometricdata}

\begin{figure}[h!]
    \centering
    \includegraphics[width=0.48\textwidth]{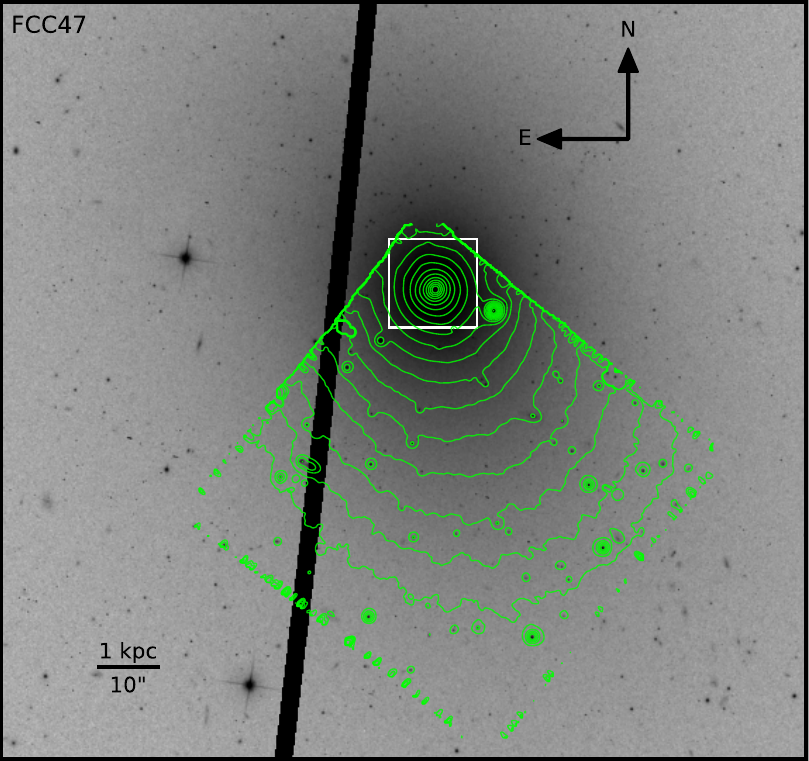}
    \caption{HST/ACS image of FCC~47 overlaid with surface brightness contours from the collapsed MUSE data cube. The white rectangle indicates a $\sim15'' \times 15''$ region centred on the galaxy nucleus, which is used to extract the stellar kinematics.}
    \label{fig:fcc47_image}
\end{figure}

We adopted the stellar mass model from \cite{Thater2023}, who derived it from high-resolution imaging taken with the Advanced Camera for Surveys (ACS) on the Hubble Space Telescope (HST). FCC~47 was observed in the F850LP filter (SDSS $z-$band) as part of the ACSFCS survey \citep[PI: Jordán,][]{Jordan2007}, see Fig.~\ref{fig:fcc47_image}. The imaging data, retrieved from the ESA Hubble Science Archive, spans a field of view (FOV) of $202'' \times 202''$ with a spatial sampling of 0\farcs05 per pixel and a spatial resolution of approximately 0\farcs09 (FWHM). The high resolution is particularly important for recovering the light distribution of the NSC. The wide FOV ensures sufficient spatial coverage. To generate mass maps, \cite{Thater2023} rescaled and multiplied the HST image with a $M_*/L$ map derived from stellar population properties, taking into account metallicity, age, and a variable IMF. These properties were obtained from the same MUSE observations used throughout this work, applying the methods described in \cite{Navarro2019} by fitting stellar population models via spectral index fitting and the penalized Pixel-Fitting (\textsc{pPXF}) technique \citep{Cappellari2004}. This approach goes beyond traditional methods based solely on surface brightness distributions by incorporating spatial variations in the stellar populations, a free IMF and their effect on the $M_*/L$. Based on the mass maps, \cite{Thater2023} constructed a multi-Gaussian expansion (MGE) model \citep{Cappellari2002}, which we re-used for our dynamical models.

\subsection{MUSE spectroscopic data}

We used MUSE/VLT observations, which were obtained during the Science Verification phase (60.A-9192, P.I. Fahrion) in September 2017 after the commissioning of the Ground Atmospheric Layer Adaptive Corrector for Spectroscopic Imaging (GALACSI) AO system \citep{Stuik2006}. It operates in the visible up to the infrared wavelength range (4650--9300 \AA). The data have a wide FOV ($60 ''\times 60''$) combined with high spatial (0\farcs7 at FWHM) and medium spectral resolution ($\Delta \lambda = 2.5$ Å FWHM at 7000 Å). We note that the spatial resolution (FWHM) is comparable to the NSC's effective radius (0\farcs75). Further details on the observations and data reduction can be found in \cite{Fahrion2019}. 

\section{Kinematics}\label{section3}

The \textsc{Bayes-LOSVD} method developed by \cite{Falcon-Barroso2020}\footnote{\href{https://github.com/jfalconbarroso/BAYES-LOSVD}{https://github.com/jfalconbarroso/\textsc{Bayes-LOSVD}}} is a Bayesian approach and represents the LOSVD as a sequence of weights applied to template spectra. This is particularly advantageous for systems that consist of multiple dynamical components or kinematically decoupled cores. The NSC in FCC~47 has more intricate and overlapping kinematic features, which make non-parametric methods like \textsc{Bayes-LOSVD} better suited to capture these details and reveal complex LOSVD structures, such as triple-peaked distributions. In contrast, parametric LOSVD models that impose constraints on the shape of the LOSVD such as Gauss-Hermite expansions \citep{vdMarel1993, Gerhard1993} may smooth out kinematic details, leading to biased dynamical inferences \citep{Reiter2025}. For computational efficiency, we created a \(\sim 15'' \times 15'' \) cutout around the central region, shown in Fig.~\ref{fig:fcc47_image}. This excluded a bright foreground star located close to the nucleus and thus removed any contamination from its light. The wavelength range from 4750 to 5450 \AA ~provides several strong absorption lines, including $H\beta$, Mg and two Fe features. This region is free from significant sky contamination, eliminating the need for masking during kinematic analysis. To facilitate transparency and reproducibility of our analysis, we report all parameters used in the extraction. We adopted 60~\kms wide velocity intervals across $\pm 700$~\kms (23 velocity intervals) and targeted a S/N of 100 (min S/N = 1 per spatial pixel). The Voronoi binning was performed internally within the \textsc{Bayes-LOSVD} code, resulting in 828 bins \citep{Cappellari2003AdaptiveTessellations}. We used the MILES stellar library \citep{MILES, Falcon-Barroso2011}, and a polynomial order of 1. We modelled the LOSVDs in all bins independently as \( \text{LOSVD}_i \sim \mathcal{N}(0, \sigma^2) \), following a normal distribution with a mean of zero and a variance \( \sigma^2 \). This assumes equal uncertainty for all LOSVDs. \textsc{Bayes-LOSVD} addresses the large number of stellar templates by applying Principal Component Analysis (PCA). We reduced the dimensionality to 5 components while preserving 99.5\% of the variance, improving efficiency without loss of accuracy.

\begin{figure*}[h!]
    \centering
    \includegraphics[width=0.9\textwidth]{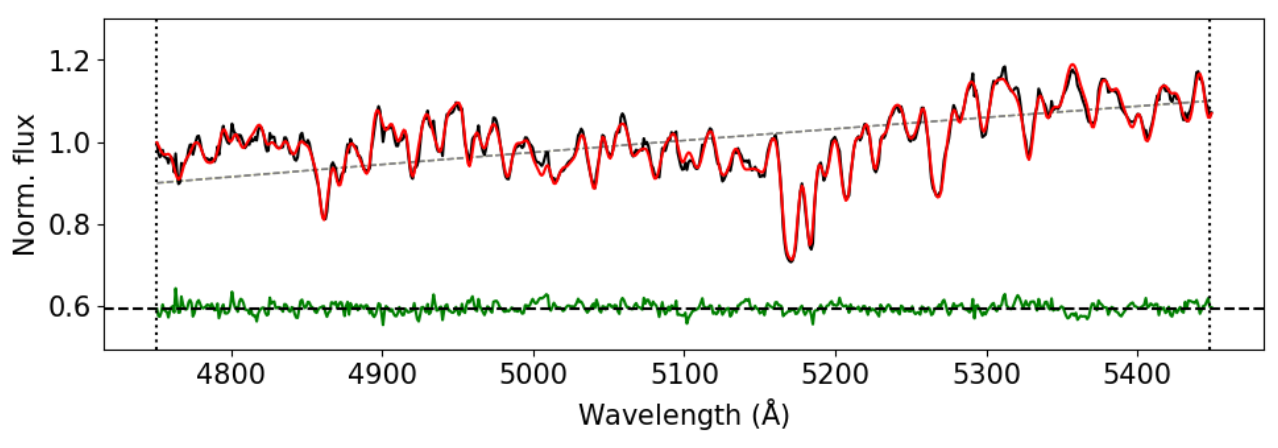}
    \caption{Spectrum fitted with \textsc{Bayes-LOSVD}. The x-axis represents the rest-frame wavelength in \AA. The y-axis shows the normalized flux. The modelled spectrum (red) closely matches the observed spectrum (black) with minimal residuals (green). Dashed lines show the continuum level of the spectrum and residual, and dotted vertical lines enclose the fitted wavelength region.}
    \label{fig:spectrum_fit}
\end{figure*}

We show an example of the \textsc{Bayes-LOSVD} fit and the observed spectrum in Fig.~\ref{fig:spectrum_fit}. It shows absorption features characteristic of a several gigayears old stellar population. We highlight the effectiveness of the fitting process as the modelled spectrum closely matches the observed spectrum with minimal residuals. We excluded only 2 of the 828 bins due to poor convergence. In addition, uncertainties in 12 bins that lie at the edge of the FOV were manually set to 10.000 to ensure only well-converged bins are considered in the dynamical models.

\begin{figure}[h!]
    \centering
    \includegraphics[width=0.5\textwidth]{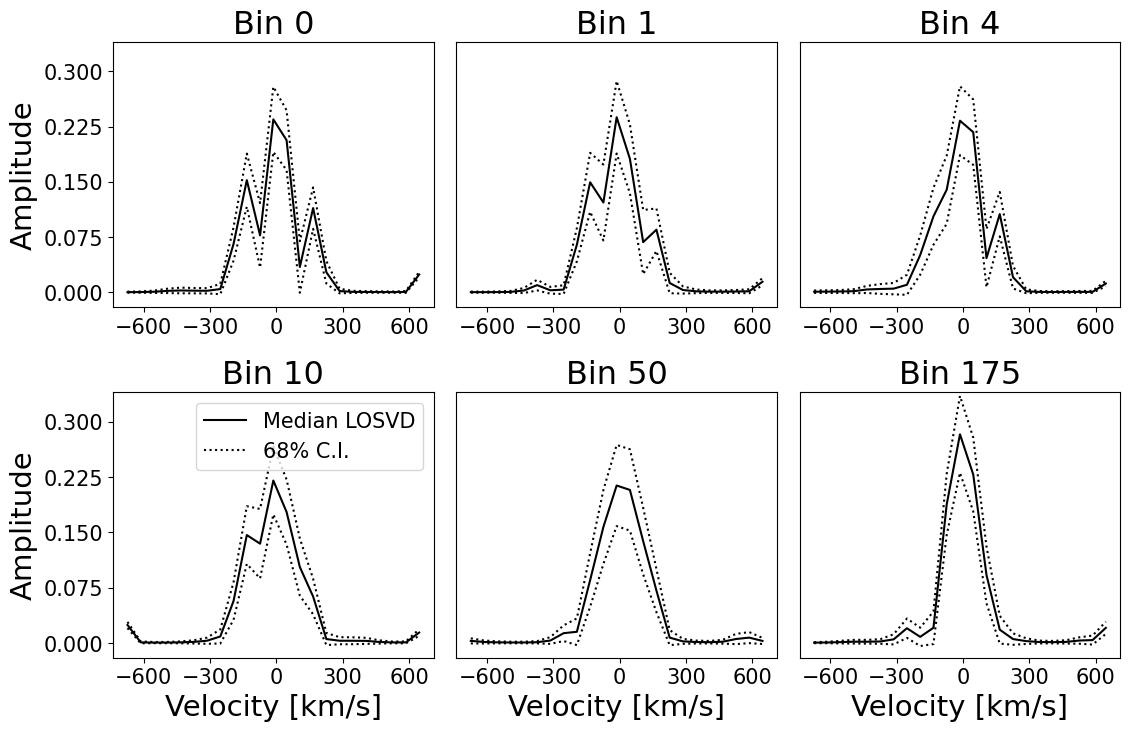}
\caption{LOSVDs extracted with \textsc{Bayes-LOSVD}. Solid lines denote the median LOSVD, the dotted lines the 68\% confidence intervals. Each panel shows a different spatial bin of the MUSE data. Bins 0, 1, 4 and 10 are likely dominated by the NSC based on their central location and kinematic features. The position of each bin in the field is shown in Fig.~\ref{fig:chi2_map}.}

    \label{fig:bayes_kinematics}
\end{figure}

We tested various regularisation methods implemented in \textsc{Bayes-LOSVD}, but found that they tend to smooth out the finer details of the kinematic structure, consistent with the findings by \citet{Reiter2025}. In their Appendix D, they show that regularisation schemes such as auto-regressive or spline priors produce smoother LOSVDs with smaller uncertainties, whereas the non-regularised setup allows the recovery of more complex, multi-peaked velocity structures, particularly in regions with intricate kinematics, at the cost of larger uncertainties. However, multiple peaks can also be artifacts due to noise, especially if no regularisation is applied. We assume that the bimodal (or even triple) peaks are genuine features of the LOSVD, but cannot fully exclude that it may still be influenced by noise. Further discussion on assessing the reality of multi-peaked LOSVDs in these small scales can be found in Appendix~\ref{Model_vs_Data}. As bin number 0 has the highest S/N, it aligns with the galaxy's nucleus. The central bins within a radius corresponding to the NSC's effective radius (see Table~\ref{tab:properties}) can be considered as those that are most likely dominated by the NSC. This corresponds to approximately the four central bins, given the MUSE pixel scale of $0\farcs2$ per pixel and the seeing conditions that also need to be taken into account ($0\farcs7$). In Fig.~\ref{fig:bayes_kinematics} we show the retrieved LOSVDs in a selection of bins. The 68\% confidence interval corresponds to the range of velocities around the median where 68\% of the total probability is concentrated. The bins that we consider NSC bins (bin 0, 1, 4, 10), where the kinematically decoupled core is situated, show several peaks in their LOSVD. Further away from the centre, the LOSVDs (e.g. in bin 50, 175) have only a single peak. More LOSVDs can be found in Appendix~\ref{Model_vs_Data}.

\section{Dynamical modelling}\label{section4}
\subsection{Schwarzschild orbit superposition models}

\begin{figure}[h!]
    \centering
    \includegraphics[width=0.5\textwidth]{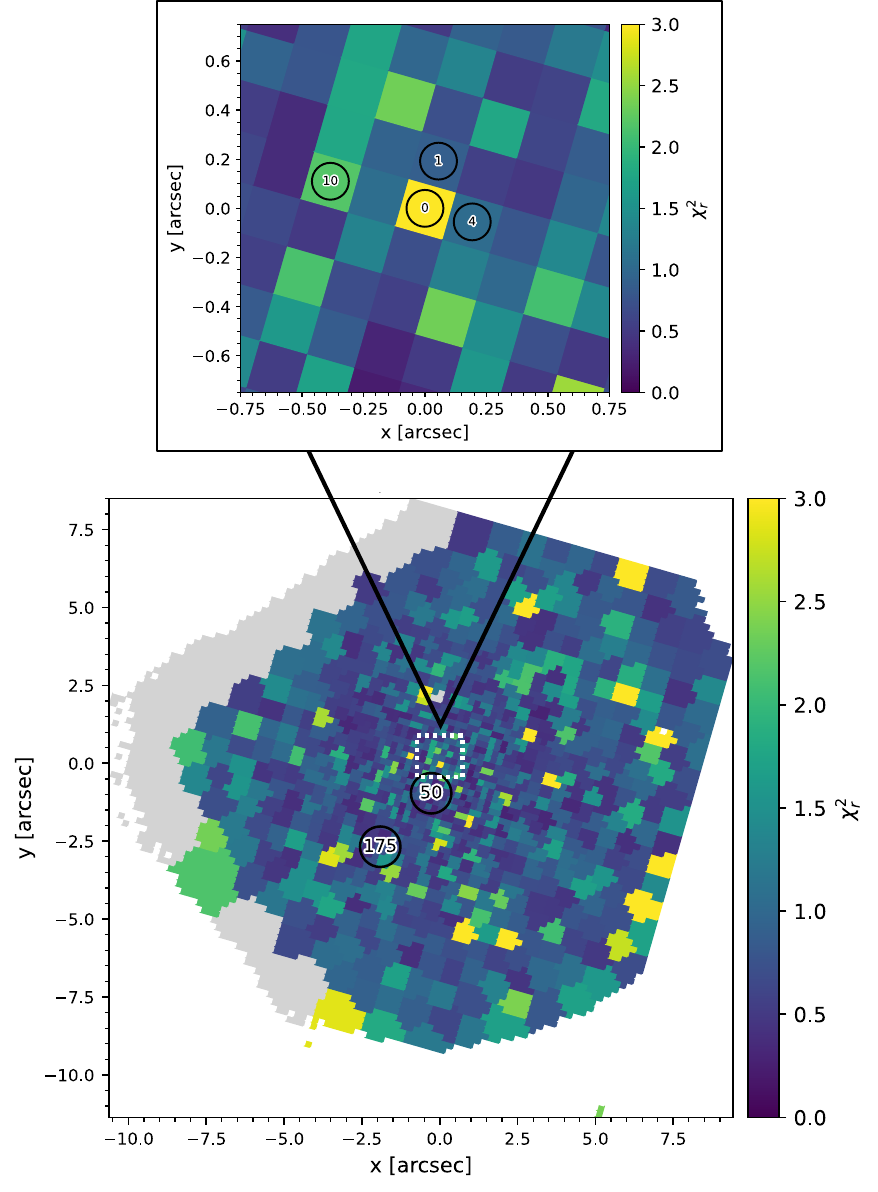}
\caption{Reduced $\chi^2$ map of the explored parameter space. The colour scale indicates the relative goodness of the fit, with darker shades corresponding to lower $\chi^2$ values.
    \textbf{Bottom:} Reduced $\chi^2$ map over the entire \(\sim 15'' \times 15'' \) field. Masked bins are coloured grey. We highlight the position of the spatial bins 50 and 175, whose LOSVD histograms are shown in Fig.~\ref{fig:bayes_kinematics}.
    \textbf{Top:} Zoom-in into the nuclear region. We highlight the position of the spatial bins 0,1,4 and 10, whose LOSVD histograms are shown in Fig.~\ref{fig:bayes_kinematics}.}
    \label{fig:chi2_map}
\end{figure}

To explore the internal orbital structure of FCC~47 and its NSC, we used the Schwarzschild orbit-superposition method \citep{Schwarzschild1979} and code based on \cite{vdBosch2008}. This technique models a galaxy by populating a triaxial gravitational potential with a representative library of stellar orbits, each integrated in the three-dimensional potential. It fits a linear combination of orbital weights that is optimised to reproduce both the observed stellar mass distribution (see Sect.~\ref{sec:photometricdata}) and the line-of-sight kinematics (see Sect.~\ref{section3}).
We utilised the triaxial DYnamics, Age, and Metallicity Indicators Tracing Evolution code (DYNAMITE v4.3.0\footnote{\href{https://github.com/dynamics-of-stellar-systems/dynamite}{https://github.com/dynamics-of-stellar-systems/\textsc{dynamite}}}; \citealt{dynamite,Thater2022}), which is particularly used for early-type galaxies, as they typically lack strongly varying stellar populations or structural features like bars or spiral arms. The total gravitational potential of a galaxy in our model comprises the stellar components, the dark matter component, and the potential due to the central SMBH. The stellar gravitational potential is described by the two-dimensional MGE (see Sec.~\ref{sec:photometricdata}), which can be directly deprojected into a three-dimensional mass density using three viewing angles $(\theta, \phi, \Psi)$ \citep{vdBosch2008}, where $\theta$ corresponds to the inclination. The viewing angles are related to the intrinsic shape parameters of the system: $p$ (intrinsic intermediate-to-major axis ratio), $q$ (intrinsic minor-to-major axis ratio), and $u$ (ratio between the projected and intrinsic major axis). The central gravitational potential is described by a Plummer potential, which includes the black hole mass $M_{BH}$ and a softening length $ a = 10^{-3}$\arcsec~ to avoid divergence. The dark matter component is parametrised as a spherical halo with a Navarro-Frenk-White proﬁle (NFW; \citealt{Navarro1996}). The total gravitational potential of all components is traced by the stellar kinematics, as stars serve as dynamical tracers that are accelerated by the total mass distribution. The BH dominates the mass distribution within its sphere of influence. We did not attempt to constrain the dark mass given the small spatial extent of the cropped MUSE field, but used the method of abundance matching by \cite{Moster2013}. We therefore fixed the concentration parameter to $C=6.4$, while the dark matter fraction, expressed as $M_{200}/M_{\star}$, was set to $1.8$. To construct the orbit library, we evenly sampled the space of three integrals of motion. Orbital energy is sampled at $n_E=35$ logarithmically spaced starting points linked to the radius of the orbits. For the second integral of motion $L_z$, the vertical component of the angular momentum, we use 11 starting points for sampling as we do for the third integral. This results in an orbit library with $35 \times 11 \times 11$ combinations of ($n_E$, $n_{I2}$, $n_{I3}$), which are mirrored for a second orbit library to produce pro- and retrograde tube orbits. A third orbit library is characterised by the energy and two spherical angles ($\theta, \phi$) for the box orbits with the same number of combinations as the tube orbits. In addition, we applied dithering by slightly perturbing the initial conditions of each orbit to achieve a smoother phase-space sampling. The resulting trajectories were co-added to form orbit bundles, each comprising $3^3 = 27$ dithered orbits in the final library. This results in $343,035$ orbits in total, or $12,705$ so-called orbit bundles. The logarithmic minimum and maximum orbit radii are defined as $\log r_{\text{min}}\text{[\arcsec]} = -2.0$ and $\log r_{\text{max}}\text{[\arcsec]} = 2.2$. The inner radius is at least 20 times (-1.3 orders of magnitude) smaller than the pixel size of the MUSE data, while the outer radius is $\approx 793$ times (2.9 orders of magnitude) larger than the MUSE pixel size. Each orbit is integrated for $200$ orbital periods, with $50,000$ sampling points per orbit. To determine the optimal orbital weights that reproduce the input data, we employ the Non-Negative Least Squares (NNLS) solver. For more information about the implementation with \textsc{Bayes-LOSVD}, see \cite{Reiter2025}. We explored the parameter space of the intrinsic shape parameters with an initial step size of 0.05 to avoid local minima, while fixing the $M_*/L$ and the BH mass to previous results of \cite{Thater2023}. Then we switched to progressively smaller step sizes around the current best-fit model, now also freeing $M_{BH}$ and $M_*/L$ (with a minimum step size of 0.01 for the intrinsic shape parameters and $M_*/L$, and 0.1 for $M_{BH}$). In this context, one iteration denotes the placement of a smaller grid around the current best fit and testing a new batch of models with finer spacing down to the minimum step size and then using the best result to recentre the next round. We repeated this process up to 31 times and evaluated $1,481$ models in total to ensure a thorough exploration of the parameter space. The derived best-fit Schwarzschild model achieved a $\chi^2_r = 0.95$, indicating good agreement between the model and the observed LOSVD, see the $\chi^2_r$-Map Figure~\ref{fig:chi2_map}. Our MUSE data does not resolve the sphere of influence of the BH, which limits our ability to obtain a robust BH mass measurement. However, we measure a $M_{\mathrm{BH}} = 6.2^{+2.6}_{-4.5} \times 10^7 M_{\odot}$ using $3\sigma$ confidence intervals, which is in agreement with prior findings by \cite{Thater2023}, who included higher spatial resolution SINFONI data and measured $M_{\mathrm{BH}} = 4.4^{+1.2}_{-2.1} \times 10^{7} M_{\odot}$ using $3\sigma$ confidence intervals. We show the best-fit $\chi^2$ distribution in the Appendix Figure~\ref{fig:kinchi2_plot}. We list the best fit model parameters in Table~\ref{tab:best-fit-parameters}.

\begin{table}[h!]
    \centering
        \caption{Best-fit model parameters with 3$\sigma$ confidence intervals.}
    \begin{tabular}{|c|c|c|c|c|c|}
        \hline
        Parameter & Best Fit & $3\sigma$ & Unit \\
        \hline
        $M/L$ & $0.64$ & $^{+0.11}_{-0.04}$ & $ M_\odot/L_\odot$ \\
        $M_{\mathrm{BH}}$ & $6.31$  & $^{+2.60}_{-4.53}$ & $\times 10^7 M_\odot$ \\
        $q$ & $0.40$  & $^{+0.00}_{-0.30}$ & --- \\
        $p$ & $0.80$ & $^{+0.10}_{-0.30}$ & --- \\
        $u$ & $0.825$  & $^{+0.17}_{-0.03}$ & --- \\
        \hline
    \end{tabular}
    \label{tab:best-fit-parameters}
\end{table}
\renewcommand{\arraystretch}{1.0}

\subsection{Best-fit orbit distribution}\label{section_bf}

Once we determined the best-fit orbit library, we analysed the orbital content by classifying the orbits according to their intrinsic angular momentum. Specifically, we computed the circularity parameter $\lambda_z = L_z / L_\mathrm{circ}(E)$, where $L_z$ is the $z$-component of the angular momentum and $L_\mathrm{circ}(E)$ is the angular momentum of a circular orbit at the same binding energy. The classification of orbits into cold, warm, hot, and counter-rotating components is central to our decomposition. The definition of the cuts is not universally fixed in the literature, and different authors have adopted slightly different schemes depending on the galaxy type and the physical scale probed. For example, \cite{Santucci2022} used thresholds optimized for massive galaxies in the SAMI survey, while \cite{Zhu2018c} applied slightly different boundaries to spirals. Nevertheless, \cite{Zhu2018a,Zhu2018b} showed for the CALIFA sample that the same scheme can be applied across the full Hubble sequence. Following \cite{Santucci2022, Zhu2018a, Zhu2018b}, we explicitly define the cuts as:

\begin{itemize}
  \item Cold orbits: $\lambda_z > 0.8$
  \item Warm orbits: $0.25 < \lambda_z \leq 0.8$
  \item Hot orbits: $|\lambda_z| \leq 0.25$
  \item Counter-rotating (CR) orbits: $\lambda_z < -0.25$
\end{itemize}

\begin{figure}[h!]
    \centering

        \includegraphics[width=0.45\textwidth]{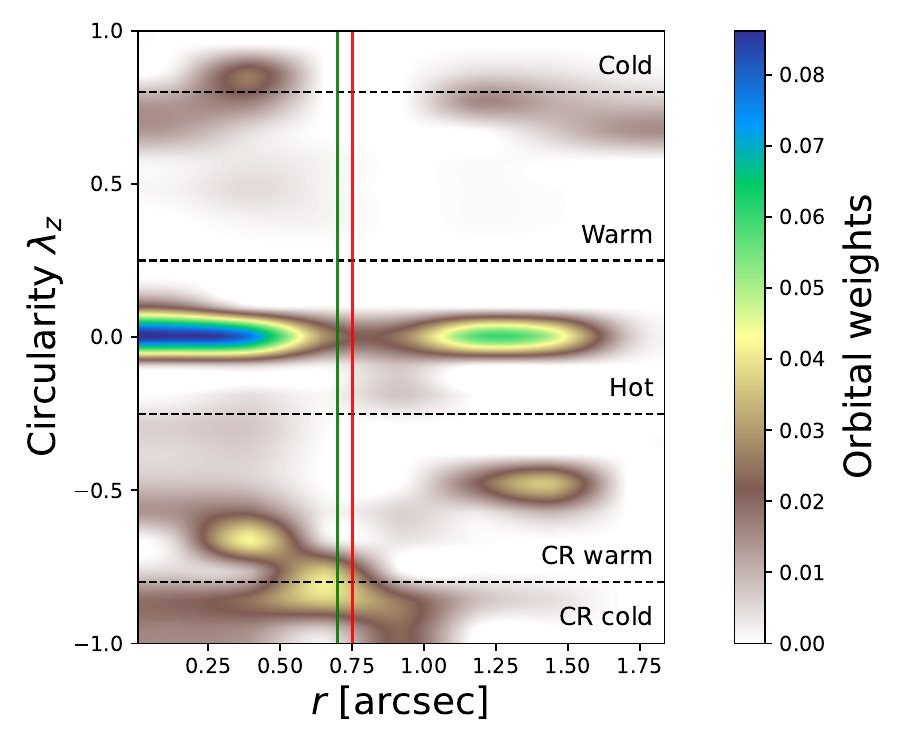}
        \label{fig:circularity_lz_1.5}

        \includegraphics[width=0.45\textwidth]{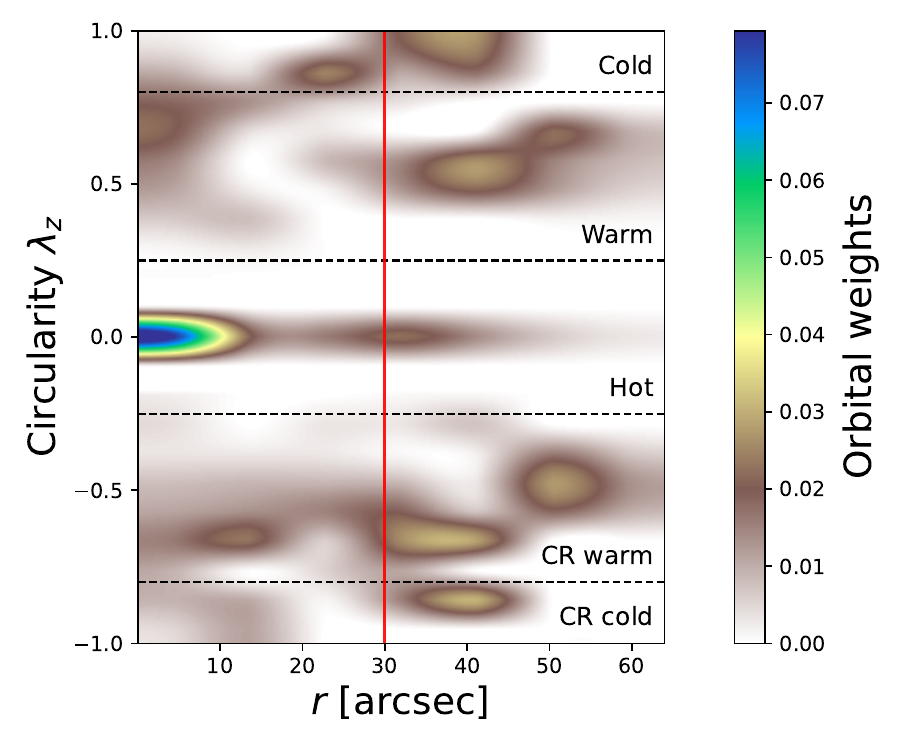}
        \label{fig:circularity_zoomout}

    \caption{Orbital distributions around the minor $z$-axis at two radial scales.
    \textbf{Top:} $r = 2''$: substantial counter-rotating motion towards the centre, the red line marks $1\,R_{\mathrm{eff,NSC}}$, the green line marks the spatial resolution at FWHM.
    \textbf{Bottom:} $r = 60''$: extended view, the red line marks $1\,R_{\mathrm{eff}}$ of the galaxy.}
    \label{fig:circularity}

\end{figure}

We experimented with slightly shifted thresholds (e.g. adopting $\lambda_z>0.7$ for the cold component or moving the hot/warm boundary to 0.3). The main decomposition remained qualitatively stable, but the relative luminosity fractions of cold versus warm orbits shifted. Importantly, the presence of the counter-rotating structures remained robust under all these variations. Therefore, the adopted thresholds are consistent with the literature and the dynamical features of FCC~47, ensuring comparability while preserving the internal kinematic substructure of the system. Figure~\ref{fig:circularity} (top panel) shows the circularity distribution in the phase space of circularity $\lambda_z$ versus intrinsic radius of the tube orbits along the minor $z$-axis within $r = 2''$. The color bar indicates the orbital weights. What immediately stands out is the significant fraction of hot orbits around $\lambda_z \approx 0$, indicative of random motion. This is a characteristic feature of early-type galaxies, and represents pressure-supported motion with little to no net angular momentum \citep{Santucci2022}. Additionally, a notable fraction of counter-rotating orbits is observed within $1''$, appearing kinematically decoupled from the rest of the system. This is consistent with previous studies, \cite{Thater2023} also found this counter-rotating core in their orbit distribution (modelling with higher spatial resolution SINFONI/VLT) and reported that the rotation originates from the inner $0\farcs4$.

\noindent Figure~\ref{fig:circularity} (bottom panel) provides a more extended view of the orbital distribution within $r\approx 60''$. The dominance of hot orbits persists until $r\approx 15''$, corresponding to approximately 0.5 $R_{\text{eff}}$. Within this region (up to $15''$), many warm and some cold prograde orbits (rotating in the direction of the galaxy's spin) are also present. In the outer regions, up to $60''$, the orbital distribution broadens, with contributions from orbits spanning a range of circularities along the $z$-axis. Beyond 1 $R_{\text{eff}}$, the orbital structure seems to change, with an increased contribution from warm prograde orbits, which could be an outer galaxy disk, that might be another distinct dynamical component. Furthermore, we detect additional counter-rotating motion from about $30''$ onward. It is also faintly visible in \cite{Fahrion2019}, but in our case orbital weights are larger. Since our analysis focuses to the central region and the kinematic input for the dynamical modelling covers only $15''\times 15''$, this could partly also be an effect of extrapolation. Summarising, the best-fit dynamical model of FCC~47 reveals a complex orbit structure, dominated by hot box orbits, as expected for early-type galaxies \citep{Santucci2022}. Counter-rotating orbits within $1''$ suggest a KDC. The orbital structure changes beyond 1 $R_{\text{eff}}$, with a growing contribution of mildly prograde orbits, indicating an extended outer component. Unfortunately, DYNAMITE is not capable of reproducing the bimodality of the LOSVDs, we suspect this is due to the very small scales. We ensured enough coverage of the phase-space by using a large orbit library, therefore enough flexibility should be given (further discussion on data vs. model comparison can be found in Appendix~\ref{Model_vs_Data}). 

\section{Orbital decomposition}\label{section5}

\begin{figure*}
    \centering
    \includegraphics[width=0.95\linewidth, height=23.5cm]{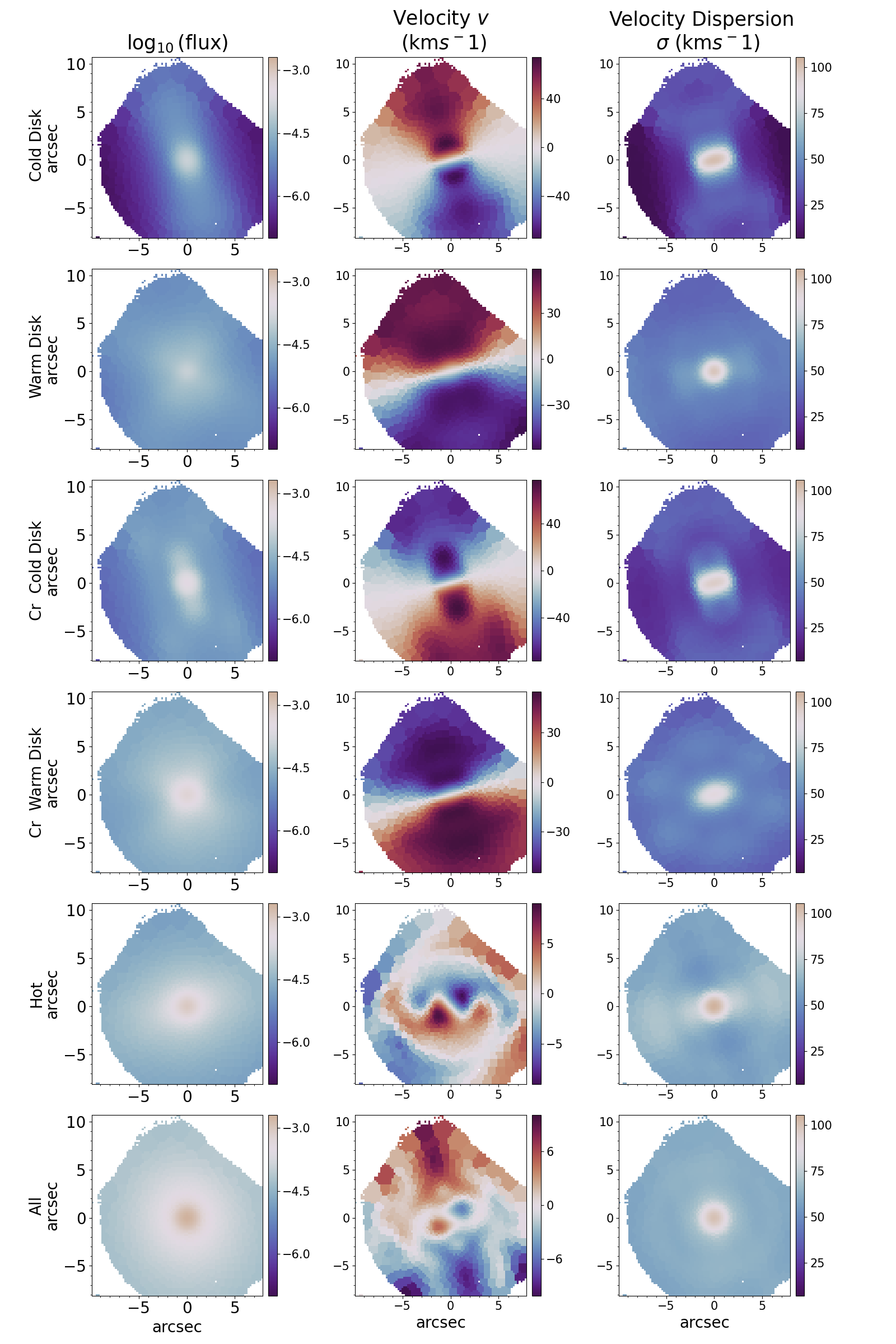}
    \caption{Orbital decomposition of the best-fit model along the minor $z-$axis based on the $\lambda_z$ circularity value. The rows correspond to the individual dynamical components, while the columns (from left to right) represent the flux contribution, velocity [km/s] and velocity dispersion [km/s] derived from the mean and the sigma of the LOSVD histograms. }
    \label{fig:decomp_lz}
\end{figure*}

The final step is the orbital decomposition of FCC~47 into dynamically distinct components. Our approach is motivated by earlier decomposition schemes (e.g. \citealt{Zhu2018a,Zhu2018b,Santucci2022}) but represents a novel application to an NSC. Unlike in massive galaxies, where decomposition typically highlights large-scale disks and bulges, here we resolve the orbital make-up of the nucleus itself, something not yet attempted at this level of detail. This represents a new and exciting way to dissect NSCs dynamically. However, the dynamical components are not strictly confined within the NSC’s effective radius as DYNAMITE operates with an average orbit radius rather than strictly confined spatial boundaries, which complicates the definition of a dynamically isolated NSC under current modelling approaches. The decomposition directly follows the cuts in the circularity parameter $\lambda_z$ (see Sect. \ref{section_bf}). After exclusion of orbits with zero weight, we use the LOSVD histograms to calculate the kinematic properties (e.g. $v, \sigma$) of the individual components, including their flux contribution to a spatial extent of $15'' \times 15''$, as shown in Fig.~\ref{fig:decomp_lz}. Because our cuts are fully explicit, this decomposition can be replicated in future studies on other galaxies with similar modelling frameworks. It is important to note that these orbits are not confined to this spatial extent, as they do not maintain a constant radius. Instead, they contribute to the potential across the entire modelled region. 

The cold disk component is characterised by high rotation, reaching a maximum velocity of \( v_{\mathrm{max}} = 75 \,\mathrm{km\,s^{-1}} \). Despite being rotationally dominated, it also exhibits a notable central velocity dispersion of up to \( \sigma_{\mathrm{max}} = 102 \,\mathrm{km\,s^{-1}} \), which we attribute mainly to the limited spatial resolution of MUSE. Its flux contribution is centrally concentrated, accounting for approximately 8\,\% of the total luminosity. The warm disk component shows lower angular momentum and extends farther out than the cold disk. It reaches a maximum velocity of \( v_{\mathrm{max}} = 56 \,\mathrm{km\,s^{-1}} \), with a peak velocity dispersion of \( \sigma_{\mathrm{max}} = 99 \,\mathrm{km\,s^{-1}} \). It contributes about 22\,\% of the total luminosity. Combined, the cold and warm disks form a disk component with a total luminosity fraction of 30\,\%, a velocity span of \( \pm~61 \,\mathrm{km\,s^{-1}} \), and a dispersion range of 35 to 101\,km\,s\(^{-1}\). The counter-rotating components appear at negative \( L_z \), forming a distinct subpopulation. The counter-rotating warm disk component is more extended and contributes 22\,\% of the total flux. It reaches a maximum velocity of \( v_{\mathrm{max}} = 55 \,\mathrm{km\,s^{-1}} \), and has a velocity dispersion between 35 and 90\,km\,s\(^{-1} \). In contrast, the counter-rotating cold disk component is more concentrated in the nucleus, with a rotational peak of \( v_{\mathrm{max}} = 77 \,\mathrm{km\,s^{-1}} \), and a dispersion of \( \sigma_{\mathrm{max}} = 97 \,\mathrm{km\,s^{-1}} \). It contributes about 5\,\% of the total flux. Together, the counter-rotating warm and cold components form a counter-rotating disk component with a luminosity share of 27\,\%, a velocity span of \( \pm~60 \,\mathrm{km\,s^{-1}} \), and a dispersion from 36 to 92\,km\,s\(^{-1} \). The hot component stands out for its high velocity dispersion, peaking at \( \sigma_{\mathrm{max}} = 106 \,\mathrm{km\,s^{-1}} \), while its rotational motion is minimal, with a maximum velocity of \( v_{\mathrm{max}} = 9 \,\mathrm{km\,s^{-1}} \). It contributes the largest single-component flux fraction, about 43\,\%. Notably, the central region shows rotation opposite to the global motion, indicating a kinematically distinct nuclear component. The overall structure of the system reveals a dynamically rich and complex configuration. Counter-rotating orbits are centralised, and orbit cancellation effects reduce the net rotational velocity in the combined maps. The maximum velocity dispersion of the total system reaches \( \sigma_{\mathrm{max}} = 101 \,\mathrm{km\,s^{-1}} \), and the overall velocity range spans up to \( v_{\mathrm{max}} = 9 \,\mathrm{km\,s^{-1}} \). The total luminosity fraction sums to 100\,\%, confirming the integrity of the decomposition. The quoted values of $v_{\mathrm{max}}$ and $\sigma_{\mathrm{max}}$ are derived from the best-fitting model and show negligible variation across models within the $3\sigma$ confidence region. To check the robustness of our results, we selected models with the largest deviation in $M_{BH}$ while keeping the intrinsic shape parameters and $M_*/L$ fixed to the best-fit values. By decomposing the models with the highest and lowest $M_{BH}$, we found no significant deviations in the kinematics for the individual components.

\section{Discussion}\label{section6}

\subsection{Resolving the complex LOSVD}

\begin{figure}[h!]
    \centering
    \includegraphics[width=1\linewidth]{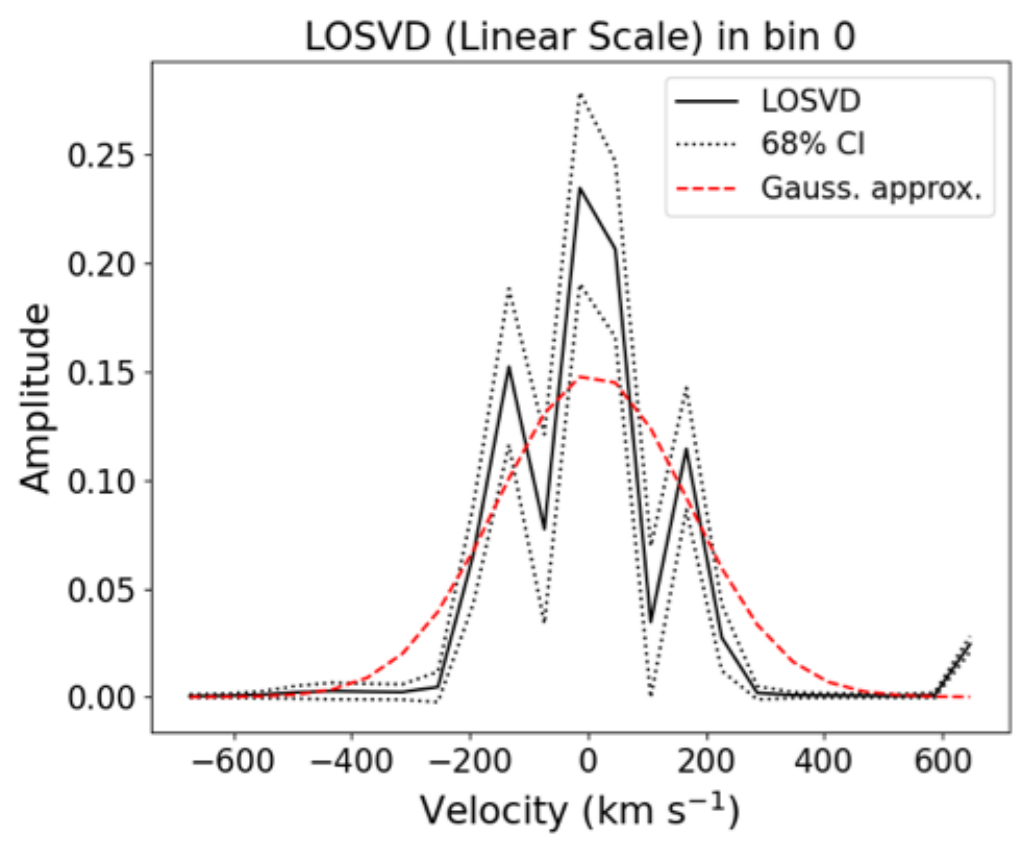}
    \caption{LOSVD of Bin 0 centred on the systemic velocity in a linear scale compared to a Gaussian approximation (in red). The dashed lines correspond to the 68\% Confidence Interval (CI). An advantage of the \textsc{Bayes-LOSVD} approach is its ability to reveal complex LOSVD structures, such as triple-peaked distributions.}
    \label{fig:losvd_comparison}
\end{figure}

We extracted the LOSVD in 828 spatial bins using \textsc{Bayes-LOSVD}. A key advantage of this approach is its ability to resolve complex LOSVD structures, including triple-peaked profiles as observed in this system. Such features can reveal substructures or kinematic components that may be smoothed over or missed entirely by traditional parametric methods. The primary peaks in the LOSVDs typically appear around $\sim 0$\,km\,s$^{-1}$, the value relative to the systemic velocity (see Fig.~\ref{fig:bayes_kinematics}). Secondary or third peaks, when present, are often offset by 200–300\,km\,s$^{-1}$. Where a single peak points to a uniform stellar population moving in a smooth velocity field (ordered manner, e.g. bin 50, 175), a double-peaked distribution indicates the presence of two distinct kinematic components, such as counter-rotating stellar populations or an embedded stellar disk (e.g. bin 4, 10). The existence of such features can also suggest interactions between different dynamical components, possibly due to the influence of the SMBH or past mergers. Triple-peaked LOSVDs, observed primarily in the innermost bins, may indicate the coexistence of more than two dynamically distinct components, reflecting the highly complex velocity structure near the nucleus (e.g. bin 0, 1). The 68\,\% confidence intervals are wider in the outskirts, likely due to lower S/N in those regions. The best-fitting orbit distribution of \citet{Thater2023}, who employed \textsc{pPXF} kinematics based on the same MUSE/VLT data together with additional SINFONI/VLT observations, shows a higher contribution of hot orbits and a lower fraction of dynamically cold orbits in the circularity distribution ($\lambda_z$). Consistently, they obtain a higher best-fit $M_*/L = 0.86 \pm 0.04$, whereas our modelling yields a lower value of $M_*/L = 0.64^{+0.06}_{-0.04}$. This coincides with the findings of \cite{Reiter2025}, indicating that fitting a parametric LOSVD with a single Gaussian component could lead to higher inferred velocity dispersions due to a tendency to fit single, smoothed profiles. We tested this by comparing the LOSVD extracted using \textsc{Bayes-LOSVD} in the central bin and a Gaussian approximation (in red) in a linear scale in Figure~\ref{fig:losvd_comparison}. The Gaussian approximation produces wider standard deviations compared to the Bayesian approach. This can propagate to overestimated dynamical masses and an inflated contribution of warm and hot orbits, characterised by higher velocity dispersion and lower angular momentum, for this system in dynamical models. In contrast, the \textsc{Bayes-LOSVD} approach captures finer kinematic details and avoids artificial smoothing, resulting in lower inferred dispersions and mass contributions, and potentially a larger fraction of cold orbits. These results have direct implications for FCC~47. By resolving multi-peaked LOSVDs, one shifts the mass budget towards a lower $M_*/L$ and therefore, hypothetically, towards a lower $M_{\text{BH}}$. Because our MUSE data do not resolve the sphere of influence of the BH, this possible decrease in $M_{\text{BH}}$ remains tentative. The higher fraction of cold, high-$\lambda_z$ orbits in our best-fit model implies a more rotation-supported nuclear structure rather than a purely pressure-supported core, consistent with a potential build-up through dissipative inflow and/or minor accretion events. Notably, \citet{Thater2023} compared their $M_{\text{BH}}$ measurement to SMBH–host scaling relations and found it to be overmassive. Introducing a spatially varying stellar $M_*/L$ map that incorporates a free IMF, in their modelling already pushed the BH mass towards the relations, but a significant offset remained. By explicitly modelling the non-Gaussian, multi-peaked LOSVDs with \textsc{Bayes-LOSVD}, our framework provides a physically motivated path to further reduce the inferred $M_{\text{BH}}$ towards the scaling relations, pending confirmation with additional high-angular-resolution data for the central region of FCC~47.

\subsection{NSC formation}

Our dynamical decomposition of FCC~47's nucleus provides constraints on the assembly history of its NSC, based solely on stellar orbital structures. Our findings confirm that the NSC in FCC~47 is characterised by high angular momentum, strong rotation, and elevated velocity dispersion, in agreement with previous studies \citep{Lyubenova2019,Fahrion2019,Thater2023}. The orbital decomposition reveals a dynamically complex structure: The cold counter-rotating disk is centrally concentrated, rotationally supported, and contributes $\sim4.8$\,\% of the total flux within our field of view. Its high angular momentum and compactness are consistent with an in situ formation scenario, likely via gas inflow during dissipative events, such as minor mergers or secular evolution. The warm counter-rotating component, contributing $\sim21.6$\,\% of the total flux, extends further out and may reflect stars formed in dynamically heated gas or radial migration of stars originally confined to the cold disk. Combined, these counter-rotating components account for $\sim26.4$\,\% of the flux in our chosen FOV and form a coherent disk-like structure embedded within the NSC. At the same time, the dominance of hot orbits — which are pressure-supported and less spatially concentrated — suggests that a significant fraction of the NSC mass formed through dissipationless processes. This is in line with the idea of star cluster migration and infall, whereby GCs or other star clusters sink to the galaxy centre via dynamical friction and dissolve, contributing stars on random orbits \citep{Guillard2016, Neumayer2020}. The overall dynamical structure of FCC~47’s NSC suggests that both accretion and dissipative processes contributed to its assembly. The strong rotation and embedded counter-rotating disk point towards a dissipative origin for at least part of the structure, while the hot, pressure-supported component likely formed through early accretion of GCs. Furthermore, the fact that the NSC rotates nearly orthogonal to the galaxy’s short axis, combined with the presence of a second kinematically decoupled disk component extending to $\sim20$\arcsec~ \citep{Fahrion2019}, supports a scenario involving a past merger that resulted in a kinematically decoupled core \citep{Lyubenova2019}.

\subsection{Linking to stellar populations}

Each of the orbital structures may carry distinct stellar population signatures, which we compare to the spectral fitting results of \citet{Fahrion2019}. FCC~47 is an overall old galaxy ($>8$ Gyr) with a central gradient towards even older ages, reaching $\sim$13 Gyr in the nucleus and NSC. \cite{Fahrion2019} found that the NSC is dominated by a super-solar metallicity population, which accounts for about 67\% of its stellar mass, alongside a secondary, more metal-poor component. Consequently, the NSC shows a super-solar metallicity in its centre that decreases to solar values at outward regions, accompanied by a high $M_*/L_g \sim 5.3$. Interestingly, the metal-rich and metal-poor subpopulations do not exhibit markedly different star formation histories (SFHs). This suggests that if the NSC (partly) formed in situ, its star formation must have been early and brief, with subsequent quenching preventing further enrichment \citep{Fahrion2019}. This cautious interpretation directly impacts the origin of the cold counter-rotating disk in our dynamical model. Its central concentration and angular momentum profile are suggestive of a dissipative origin. While such structures are commonly interpreted as the result of in situ star formation from gas inflow, the lack of distinguishable SFHs between stellar populations in FCC~47 implies that this component may have formed via early, short-lived star formation or even from stars/external star clusters brought in by a merger event and dynamically settled into a disk-like configuration. The warm counter-rotating component could represent the remnants of a heated stellar disk or stars dynamically redistributed through secular evolution. The age spread identified by \citet{Fahrion2019}, with a notable contribution from $\sim8$\,Gyr-old stars in the disk region, supports this interpretation as a result of dynamical processes acting on pre-existing stellar populations. The hot component, which dominates the mass budget and is dispersion-supported, is consistent with formation through accretion. \citet{Fahrion2019} found that FCC~47 hosts a broad, metal-poor GC population with metallicities around $-1.5$\,dex and a total mass of only $\sim1.2 \times 10^8\,M_\odot$, which is roughly 17\% of the NSC’s estimated stellar mass ($\sim7 \times 10^8\,M_\odot$). This mass discrepancy, along with the mismatch in metallicities between the GCs and the NSC, implies that while accretion likely contributed significantly to the NSC’s assembly, especially to its hot component, it cannot account for its full build-up. Additional stars must have been accreted through mergers or formed early from already enriched gas. In summary, the stellar population properties derived in \citet{Fahrion2019} map well onto the kinematic substructures identified in this work. The cold and warm counter-rotating components are dynamically distinct but do not correspond to clearly separated SFHs, suggesting an early, rapid formation scenario likely tied to merger-driven gas inflows and stellar redistribution. The hot, dispersion-supported structure is consistent with a GC-driven origin, though insufficient in mass to explain the NSC on its own. These findings collectively support a composite formation pathway involving both dissipative and dissipationless processes, in line with simulations and observational results for other early-type galaxy nuclei \citep[e.g.][]{Hartmann2011,Antonini2013,Guillard2016}.

\section{Conclusion}\label{section7}

This study presents an orbital decomposition of the nuclear regions in the early-type galaxy FCC~47, based on Schwarzschild modelling within the DYNAMITE framework and kinematics extracted using \textsc{Bayes-LOSVD}. The results lead to the following key conclusions:

\begin{itemize}
    \item The central region (i.e. the NSC) exhibits a dynamically complex structure comprising a centrally concentrated cold counter-rotating disk, a more extended warm counter-rotating component and a dominant hot, dispersion-supported component.
    \item The \textsc{Bayes-LOSVD} method, applied here for the first time to an NSC, effectively resolves asymmetric and multi-peaked LOSVDs and reveals kinematic subpopulations. It can improve the reliability of orbit decomposition and reduces artificial broadening effects that can bias velocity dispersion and mass estimates.
    \item The comparison with stellar population results from \citet{Fahrion2019} supports a composite formation scenario:
    \begin{itemize}
        \item The cold component is dynamically consistent with dissipative formation.
        \item The warm component may have formed via dynamical mixing or secular processes following a merger or gas inflow event.
        \item The hot component is best explained by early GC accretion, but the NSC’s higher mass and metallicity indicate that additional formation channels are required.
    \end{itemize}
    \item The total mass of FCC~47’s GC system appears to be only $\sim 17\%$ of the NSC’s stellar mass, and the GC metallicities might be significantly lower, reinforcing the need for an additional mechanism such as early, rapid star formation from accreted gas.
\end{itemize}

Combining dynamical modelling with spatially resolved stellar population analysis with the \textsc{Bayes-LOSVD} extraction is subject of future work to improve constraints in the $M_*/L$ variations within the NSC. This will be crucial for disentangling the contributions of different formation mechanisms and refining central BH mass estimates in dense galactic nuclei.

\begin{acknowledgements}

We thank the anonymous referee for helpful comments and suggestions that improved this manuscript. This research was supported by the MUNI Award in Science and Humanities MUNI/I/1762/2023. We gratefully acknowledge the support of the European Southern Observatory (ESO) in Garching, which hosted JL for two research stays and provided financial support through the science internship scheme. We also thank the University of Vienna for funding the research through the Short-term grant abroad (KWA), which enabled JL to work on her thesis on-site at ESO in Garching. AFK acknowledges funding from the Austrian Science Fund (FWF) [grant DOI 10.55776/ESP542]. KF acknowledges funding from the European Union’s Horizon 2020 research and innovation programme under the Marie Sk\l{}odowska-Curie grant agreement No 101103830. JF-B acknowledges support from the PID2022-140869NB-I00 grant from the Spanish Ministry of Science and Innovation. IB has received funding from the European Union’s Horizon 2020 research and innovation programme under the Marie Sk\l{}odowska-Curie grant agreement No 101059532, this project was extended for 6 months by the Franziska Seidl Funding programme of the University of Vienna. This work is based on observations collected at the European Organization for Astronomical Research in the Southern Hemisphere under ESO programme 60.A-9192. 

The computational results presented have been achieved using the Vienna Scientific Cluster (VSC-4 and VSC-5). This research made use of \textsc{SciPy} \citep{Virtanen2020} and \textsc{Stan} \citep{Stan}.
\end{acknowledgements}

\bibliographystyle{aa}   
\bibliography{References}

\appendix
\begin{appendix}
\onecolumn

\section{Testing model reliability} \label{Model_vs_Data}

\subsection{Multiple features of the central \textsc{Bayes-LOSVDs}}

To verify that the multiple peaks seen in the \textsc{Bayes-LOSVDs} are not artifacts due to noise, especially since no regularisation was applied, we directly compare LOSVDs across several neighbouring bins, see Fig.~\ref{fig:kinematic_map}. If the number and positions of peaks build a pattern between adjacent apertures, this supports that these features are real. Conversely, random fluctuations in the noise would lead to peaks shifting unpredictably between bins. Another potential source for apparent substructure could be sky-line residuals in the spectra; this is unlikely here since the LOSVDs were extracted from the optical spectral range (covering the Mg\,b, Fe\,I and H$\beta$ regions) rather than the red end, where the Ca\,triplet lines are, and where the sky residuals may contaminate the spectrum. 

A closer inspection of the central \textsc{Bayes-LOSVDs} reveals repeatable substructure. Several neighbouring apertures show two or sometimes three distinct peaks that are roughly symmetric around $v \approx 0~\mathrm{km\,s^{-1}}$, with similar velocity separations between peaks (typically $\approx \pm100$--$200~\mathrm{km\,s^{-1}}$) and comparable amplitudes and peak widths across adjacent bins. This spatial coherence is particularly evident in the central region, where, for example, bins 0, 1, 6 and 26 display similar aligned peaks of comparable strength. While we cannot fully exclude that part of the observed multi-peaked structure may still be influenced by noise, the repeated and symmetric pattern across neighbouring apertures argues against random fluctuations as the sole cause. The overall consistency suggests that these multiple peaks are genuine features of the NSC.

\subsection{Reproducing features in dynamical models}

\begin{figure}[h!]
    \centering
    \includegraphics[width=1\textwidth]{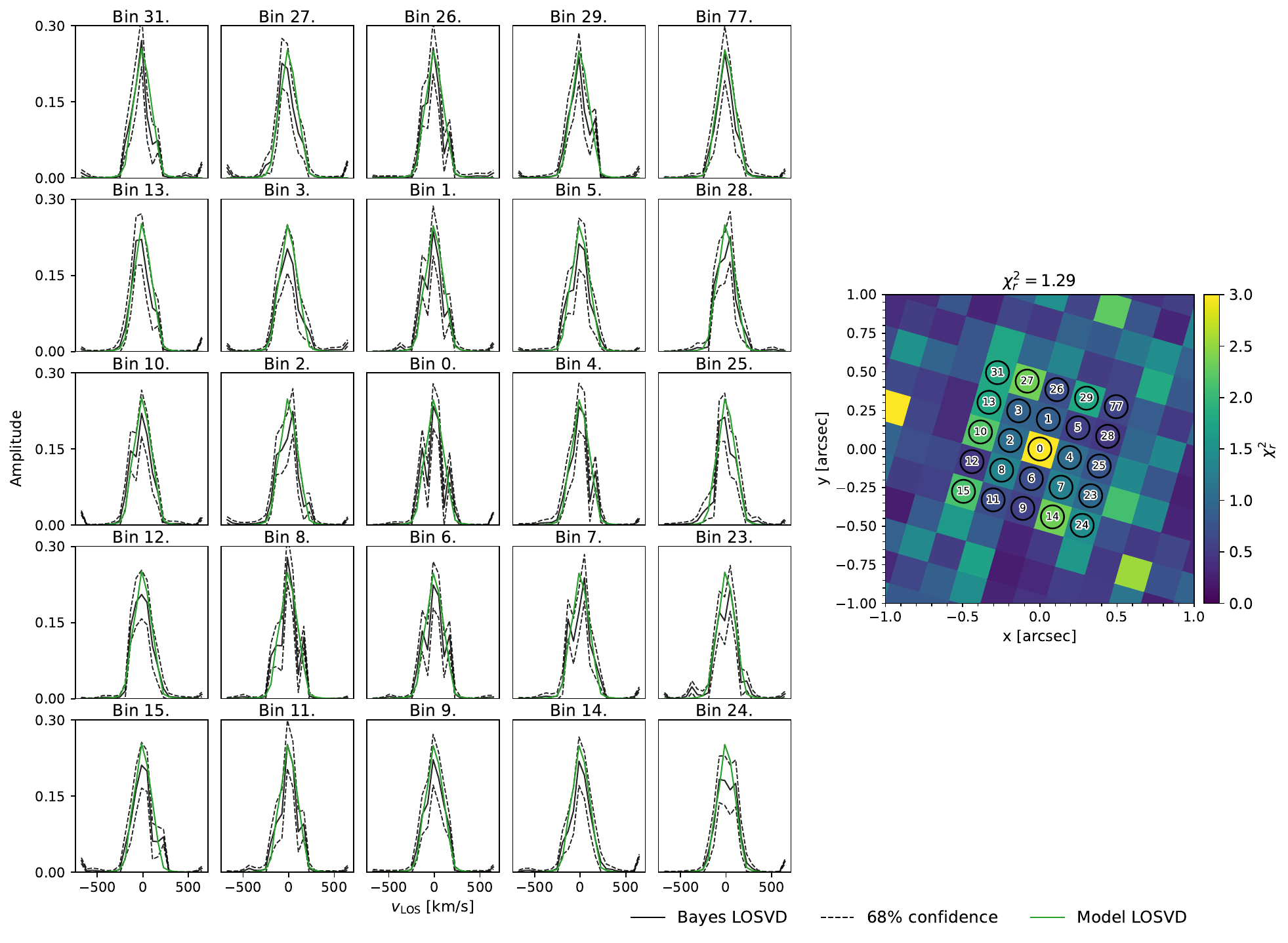}
\caption{
    \textbf{Left:} Comparison between the measured Bayes-LOSVDs (in black) and the corresponding weighted model LOSVDs from \textsc{DYNAMITE} (in green) for the central $5\times5$ spatial bins. The dashed black lines display the 68\% confidence interval. 
    \textbf{Right:} Reduced $\chi^2$ map of the nuclear region, highlighting the bins that are the most likely affected or dominated by the NSC.}
    \label{fig:kinematic_map}
\end{figure}

\begin{figure}[h!]
    \centering
    \includegraphics[width=1\textwidth]{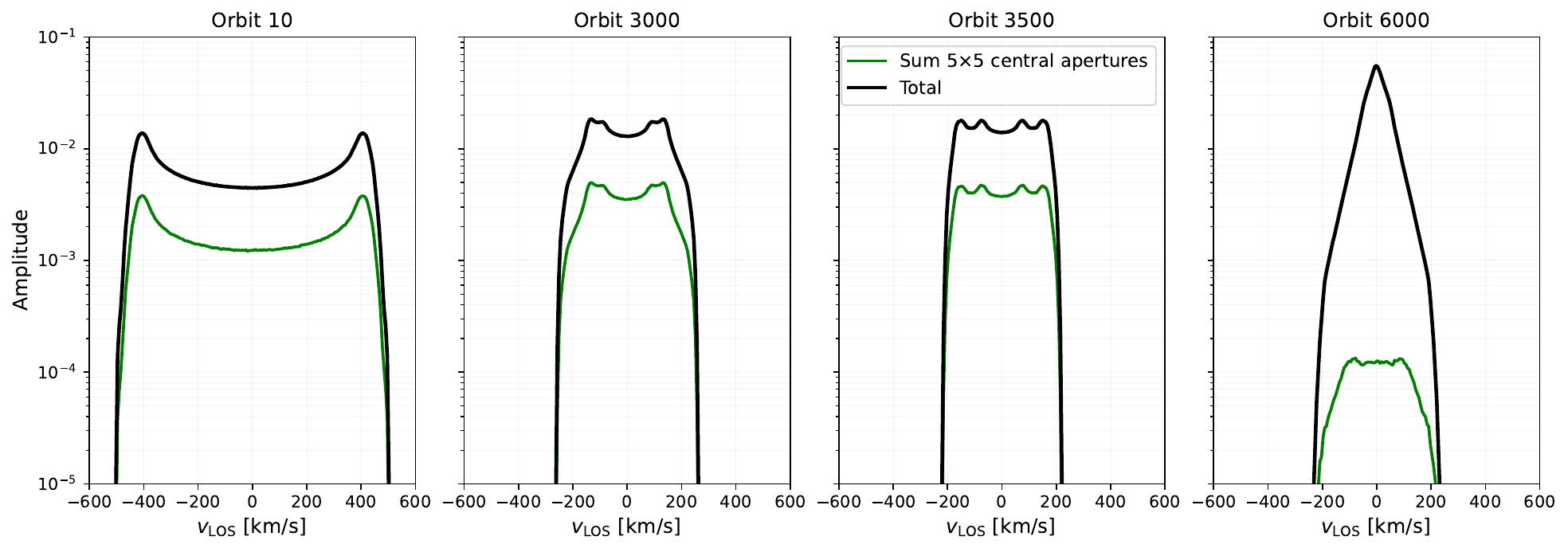}
 \caption{
        Example LOSVDs of four representative orbits from the dynamical model. 
        The black curves show the LOSVD of each orbit across all apertures, while the green curves correspond to the summed contribution of the central $5\times5$ apertures to the orbit LOSVD. Although some orbits exhibit bimodal structures, the weighted sum of the orbit LOSVDs does not reproduce the pronounced multi-peaked features observed in the \textsc{Bayes-LOSVDs}, see Fig.~\ref{fig:kinematic_map}.
    }

    \label{fig:Dyna_orbits}
\end{figure}

Furthermore, we test whether \textsc{DYNAMITE} can reproduce the observed multiple features within the dynamical models. Figure~\ref{fig:Dyna_orbits} shows the LOSVDs of four exemplary orbits, with the black line showing the total LOSVD of all spatial apertures and in green the sum of the contributions of the central $5\times5$ apertures. We show that individual orbits in the model can exhibit multimodal LOSVDs. However, \textsc{DYNAMITE} assigns each orbit a weight based on how well it reproduces the observed kinematics, thereby orbits with strongly mismatched LOSVDs receive negligible weight and thus contribute little to zero to the total model LOSVD. For a given aperture, the model LOSVD (in Fig.~\ref{fig:kinematic_map}) represents the weighted sum of all orbit LOSVDs that contribute to that spatial bin. Additionally we show the reduced $\chi^2$ map and  highlight the central bins, that are affected or likely dominated by the NSC. The reduced $\chi^2_r$ in the nuclear region ($\chi^2_r = 1.29$) is slightly higher than the value obtained over the full spatial extent of the observables ($\chi^2_r = 0.95$), corresponding to the central $\sim 15'' \times 15''$ cutout. This indicates marginally poorer fits in the nuclear region relative to the field-averaged value, reflecting the increased complexity of the nuclear kinematics. In our case, while several orbits show bimodal or even multi-peaked LOSVDs, the model does not reproduce the multiple peaks seen in the \textsc{Bayes-LOSVDs} (Fig.~\ref{fig:kinematic_map}). This suggests that the model lacks sufficient flexibility to fit this level of small-scale detail, even though in principle the method is capable of doing so.

We used an orbit library with a sampling of $(n_E, n_{L_z}, n_I) = (35,\,11,\,11)$ and dithering $=3$, therefore providing adequate phase-space coverage. For comparison, \citet{Reiter2025} successfully reproduced LOSVD bimodality in a counter-rotating system using a smaller orbit library of $(21,\,10,\,7)$. However, in that case, the bimodality arises from two larger-scale counter-rotating components, whereas in FCC~047 the features occur on much smaller scales within the NSC. Hence, despite the overall good fit (as indicated by the low $\chi^2$ values), the model does not capture this fine structure—most likely due to the limited spatial resolution and the intrinsically small spatial scale of the NSC.

\clearpage

\section{\textsc{DYNAMITE} $\chi^2$ distribution} \label{Kinchi2_plot}

\begin{figure}[h!]
    \centering
    \includegraphics[width=1\textwidth]{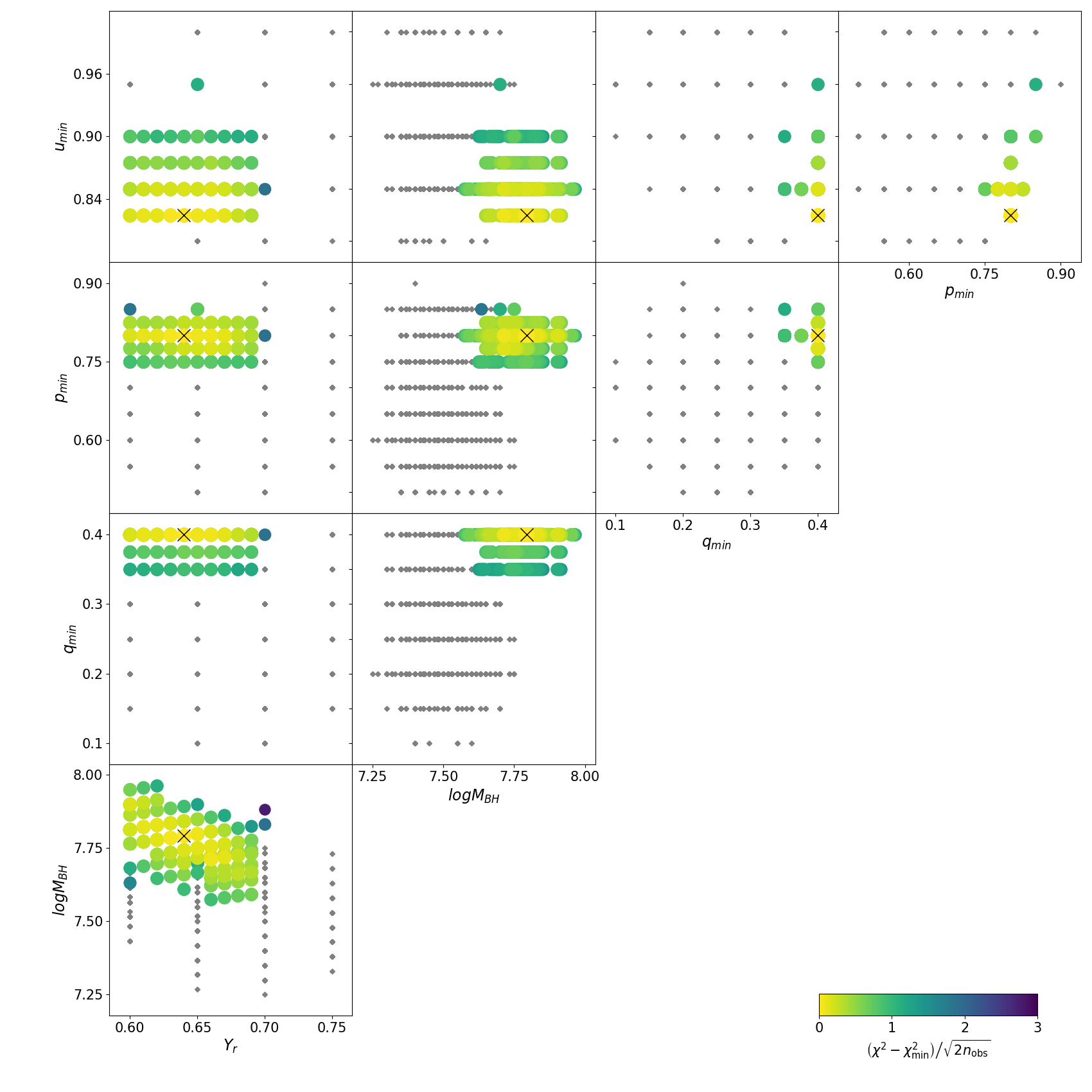}
\caption{$\chi^2$ distribution: Each point denotes one model, colour-coded by the normalized $\Delta\chi^2 = (\chi^2 - \chi^2_{\mathrm{min}}) / \sqrt{2n_{\mathrm{obs}}}$ as indicated by the colourbar. $Y_r$ is the dynamical $M_*/L$. The parameters shown are $Y_r$, the black hole mass $\log M_{\mathrm{BH}}$, and the intrinsic shape parameters $(q_{\mathrm{min}}, p_{\mathrm{min}}, u_{\mathrm{min}})$. The black cross marks the best-fit model corresponding to the global $\chi^2_{\mathrm{min}}$.}

    \label{fig:kinchi2_plot}
\end{figure}

\twocolumn
\end{appendix}

\end{document}